\documentclass[conference]{IEEEtran}
\pdfoutput=1
\usepackage{cite}
\usepackage{amsmath,amssymb,amsfonts}
\usepackage{graphicx}
\usepackage{caption}
\usepackage{subcaption}
\usepackage{multirow}
\usepackage{xcolor}
\usepackage{url}

\usepackage{pifont}%
\newcommand{\cmark}{\ding{51}}%
\newcommand{\xmark}{\ding{55}}%

\begin{document}

\title{{\normalsize\normalfont\color{red} This paper appeared in the 2024 34th International Conference on Field-Programmable Logic and Applications (FPL)}\\[0.8em]
The Road Less Traveled: Congestion-Aware NoC Placement and Packet Routing for FPGAs}

\author{
\IEEEauthorblockN{Soheil Gholami Shahrouz and Vaughn Betz}
\IEEEauthorblockA{
\textit{Department of Electrical and Computer Engineering} \\
\textit{University of Toronto}, Toronto, Canada \\
s.shahrouz@mail.utoronto.ca, vaughn@eecg.utoronto.ca}
}

\maketitle

\begin{abstract}
To help scale to ever-larger and more complex designs, recent FPGA architectures now integrate network-on-chips (NoCs). NoCs help transfer high-bandwidth data over long distances within the chip without using scarce low-delay long routing wire segments. While NoC-enhanced FPGAs aid system integration and design reuse, they also complicate FPGA computer-aided design (CAD) flows by introducing new constraints and metrics. Placement and routing need to optimize NoC metrics like latency and bandwidth utilization 
and avoid link oversubscription (congestion), while simultaneously optimizing the programmable routing resource usage of the design modules attached to NoC routers.

In this work, we develop several new approaches to reduce NoC congestion while minimizing the impact on other design metrics. First, we incorporate a NoC link congestion cost into the placement engine of the open-source CAD flow, versatile place \& route (VPR).
Second, we integrate \textit{turn model} NoC routing algorithms into the placement engine to leverage path diversity to further reduce congestion. On average over a suite of 29 benchmarks, combining placement congestion modeling with turn model packet routing reduces NoC congestion by 90.7\% at the cost of increasing aggregate bandwidth demand by 4\%.
In cases where the enhanced placement engine and NoC routing fail to fully resolve congestion, we formulate NoC routing as a Boolean satisfiability (SAT) problem. This approach yields significant additional improvements; the combined algorithm reduces congestion by 95.1\% compared to the baseline placement. Finally, we enhance the reinforcement learning (RL) agent in VPR's placement engine by introducing a NoC-aware move type, resulting in an 8.8\% reduction in wirelength on designs that make extensive use of the NoC.
\end{abstract}


\fontsize{9.9}{11.0}\selectfont

\section{Introduction}\label{sec:intro}
With the unprecedented demand for computing power, domain-specific accelerators have become an indispensable part of today's computing landscape. Among various accelerators, field-programmable gate arrays (FPGAs) have proven effective in numerous data-intensive computing tasks, owing to their spatial parallelism and reconfigurability. This spatial parallelism is achieved by integrating a large number of routing wires and programmable switches to enable configurable fine-grained custom connectivity between millions of logic blocks, block RAMs (BRAMs), and digital signal processing (DSP) blocks \cite{boutros2021fpga}.

While technology scaling has increased the logic density in FPGAs, it has also led to higher wire resistance, resulting in increased signal delay per logical distance \cite{nikolic2021global}. Simultaneously, the rise of data-intensive applications such as machine learning \cite{ibrahim2023extending}, network processing \cite{wang2022fpganic}, and storage acceleration \cite{salamat2021nascent} has driven the demand for bandwidth. To address this growing demand, modern FPGA devices incorporate several complex hardened I/O interfaces and memory controllers, such as PCIe, high-bandwidth memory (HBM), and Ethernet interfaces. Hardening such interfaces addresses the requirement for transferring high-bandwidth data into and out of the FPGA.

However, moving such high-bandwidth data within the FPGA fabric places increasing demand on routing resources. This is because hardened I/O and memory interfaces are typically clocked at much higher frequencies than the FPGA fabric. When the data enters the fabric, it must be transmitted with wider buses to compensate for the lower clock frequency of the fabric, thereby increasing the demand for routing wire segments. High programmable routing resource utilization renders timing closure more challenging, necessitating several design iterations to meet timing constraints.

Various efforts have aimed to optimize bus routing in FPGAs through both architectural \cite{ye2005using} and CAD \cite{liu2019rapidroute} approaches. However, the architectural approach is not flexible enough to enable efficient routing for a wide variety of circuits, and as a result, it has not been widely adopted by commercial FPGAs. On the other hand, the CAD approach requires user intervention in the CAD flow, typically involving manual floorplanning.
Several commercial FPGA devices have taken a new approach by integrating Networks-on-Chip (NoCs) to improve FPGA data movement capabilities. For example, AMD's Versal architecture \cite{swarbrick2019network} features a hardened NoC for high-bandwidth data transfers between the FPGA fabric, DDR memory controllers, processors, and AI Engines. Achronix's Speedster7t \cite{achronix2019} includes a 2D NoC, and Intel's Agilex M device incorporates horizontal NoCs for efficient access to high-bandwidth memory and I/O interfaces. The integration of NoCs into commercial FPGAs has also shown promise in reducing compilation times and enhancing design reuse \cite{nguyen2023spades}.

Integrating NoCs into FPGAs presents a new challenge for FPGA CAD flows. CAD algorithms now need to optimize both traditional fabric metrics (e.g., total wirelength and critical path delay) and NoC-related metrics such as aggregate bandwidth utilization and traffic flow latency while ensuring NoC-related constraints are met. The first constraint is the maximum permissible latency for each traffic flow, set by the user or determined through system-level simulations \cite{boutros2022rad}. Violating these latency constraints can negatively impact system performance. Guo et al. \cite{guo2021autobridge} observed that arbitrary pipelining of some latency-insensitive interfaces may introduce latency imbalances into parallel paths, reducing overall throughput. Similarly, violating latency constraints in a NoC-mapped application can also degrade performance. The second constraint is implicit: the total allocated bandwidth traveling through NoC links must not surpass their capacity. If this happens, the latency experienced by each flit or packet increases as intermediate buffers fill up, ultimately causing backpressure at the compute units and reduced throughput.

In this work, we incorporate NoC link congestion modeling as part of the placement cost function in VPR's placement engine. When a deterministic minimal NoC routing algorithm (XY-routing \cite{seitz1988architecture}) is utilized, congestion modeling somewhat mitigates NoC link congestion but is usually insufficient to fully resolve congestion. To address this limitation, we integrate turn model minimal routing algorithms \cite{glass1992turn} into the placement engine to leverage the path diversity inherent in mesh NoC topologies. The combined approach of a congestion-aware placement cost function and path diversity-aware minimal routing algorithms effectively resolves congestion in most benchmarks, with a slight increase in aggregate bandwidth utilization. In cases where VPR's congestion-aware placement engine fails to resolve NoC congestion, we define NoC routing as a Boolean satisfiability (SAT) problem and invoke a SAT solver to alleviate congestion by rerouting some traffic flows through longer detours, thereby reducing the bandwidth utilization of congested links. The main contributions of this work are as follows:

\begin{itemize}
    \item Modeling NoC link congestion within an open-source FPGA CAD flow\footnote{The source code of this work has been integrated into the VTR project available at https://github.com/verilog-to-routing/vtr-verilog-to-routing.}.
    \item Resolving NoC congestion by simultaneously optimizing NoC placement and routing and exploiting the intrinsic path diversity in mesh topologies.
    \item Formulating NoC routing as a SAT problem to enable congestion reduction via detours.
    \item Proposing NoC-aware packing and placement optimizations for conventional FPGA fabric primitives and blocks to improve total wirelength.
    \item Experimental evaluation of the proposed algorithms and an ablation study of different optimizations.
\end{itemize}

The rest of the paper is organized as follows: Section~\ref{sec:related_work} briefly reviews previous related work. The NoC congestion model, integration of path-diverse NoC routing algorithms, and SAT formulation of NoC routing are reviewed in Section~\ref{sec:congestion_aware_noc_mapping}. Section~\ref{sec:pack_place_opt} proposes a NoC-aware packing optimization and a NoC-aware move type used in the placement stage. Section~\ref{sec:results} experimentally evaluates the proposed algorithms and optimizations using both synthetic and real benchmarks. Section~\ref{sec:conclusion} concludes the paper and proposes possible future research directions.

\section{Related Work}\label{sec:related_work}
\subsection{NoC-enhanced FPGA Architectures}
Modern FPGA devices are integrating hardened high-speed serial interfaces and memory controllers to meet the increasing demand for high-bandwidth data transfer. For instance, while the Virtex-7 690T FPGA board could only provide a peak DRAM bandwidth of 13 GB/s, the Alveo U280 board can deliver a peak HBM bandwidth of 425 GB/s \cite{choi2021hbm}. In the Alveo U280, HBM memory controllers utilize 64-bit I/O buses operating at 900 MHz (double data rate). To match the peak memory bandwidth, the memory communicates with the fabric through a 256-bit AXI interface clocked at 450 MHz. However, achieving timing closure for a 450 MHz clock frequency is challenging. As a result, the fabric logic typically uses a 512-bit interface operating at a lower frequency. Thus, to fully utilize the peak HBM bandwidth, 32 AXI interfaces with data read/write channels as wide as 512 bits need to be routed. As reported by Guo et al. \cite{guo2021autobridge}, this places significant pressure on routing resources, particularly near HBM channels. To address this issue, commercial FPGA vendors have started to integrate NoC routers into their modern devices. For instance, AMD Versal features a NoC with 128-bit links operating at 1 GHz to match the peak bandwidth of DDR3 channels \cite{swarbrick2019network}.

When targeting a NoC-enhanced FPGA, designers can instantiate \textit{logical} NoC routers and connect design modules to them through latency-insensitive interfaces like AXI \cite{amba_axi_spec} to enable transmitting high-bandwidth traffic flows. The other endpoints of these traffic flows are connected to other NoC routers. Therefore, instead of routing wide buses through the FPGA fabric to enable delivering high-bandwidth traffic flows, they are carried by NoC links.

\subsection{CAD Support for NoCs}
Given a user design with instantiated logical NoC routers, it is now the CAD flow's responsibility to map the logical NoC routers to \textit{physical} NoC routers in the target FPGA device and route the traffic flows between them. Even though the integration of hardened NoC routers has proven to be effective at reducing programmable routing congestion and improving design reusability, their integration into the FPGA CAD flow has not been studied extensively. Abdelfattah et al. \cite{abdelfattah2016lynx} introduced LYNX to map FPGA accelerators to NoC-enhanced FPGAs. They employ a simulated annealing optimizer to minimize aggregate bandwidth and latency, and forbid modules from being connected to two NoC routers so that their logic blocks are not stretched between two NoC routers. However, their method is a pre-processing step before the FPGA CAD flow as they do not consider how NoC placement impacts the placement of fabric primitives (e.g., LUTs, registers) and connectivity. AMD has revealed a high-level overview of NoC mapping and routing algorithms in their commercial CAD flow \cite{swarbrick2019network}. They have adopted a two-step approach where they first place NoC routers and route traffic flows between them. In the second stage, NoC routers are locked down, and the rest of the netlist is placed and routed. If during the second stage, it turns out the NoC mapping has made the second stage optimization too complex, they repeat the first step to find a more suitable NoC mapping. In NoC placement, they use some heuristics like netlist connectivity and traffic flow locality to find a suitable mapping. Then, NoC routing is modeled as a SAT problem to route all traffic flows while avoiding deadlock and link oversubscription. However, the details of NoC placement heuristics and the SAT formulation have not been published. Srinivasan et al. \cite{srinivasan2023placement} integrated NoC architecture modeling and placement optimization into VPR. They co-optimize NoC and conventional netlist placement by using a simulated annealing optimizer. Although they add aggregate NoC bandwidth usage and traffic flow latencies to the SA's cost function, they ignore NoC congestion. To avoid deadlock in the mapped NoC solution, they use the XY-routing algorithm which is guaranteed to be deadlock-free, but which is unable to exploit path diversity. This work builds upon \cite{srinivasan2023placement} and uses the same terminology.

\subsection{NoC Routing Algorithms}\label{sec:noc_routing}
Given a NoC topology and the traffic flows between NoC routers, NoC routing determines the links that each traffic flow should traverse to reach its destination. The development of NoC topologies, architectures, and routing algorithms has largely aimed at addressing scalability issues inherent in on-chip interconnections such as buses and crossbars. In the literature, considerable attention has been devoted to adaptive routing algorithms \cite{li2006dyxy, singh2003goal, gratz2008regional}, which dynamically adjust traffic flow routes in response to network conditions. These sophisticated algorithms are useful in multiprocessors characterized by highly variable NoC workloads and endpoints capable of handling out-of-order packet delivery.

However, in the context of NoC-enhanced FPGAs, the traffic flow pattern remains relatively constant once the FPGA is programmed. Additionally, most design modules lack support for out-of-order packet delivery. Consequently, routing algorithms employed in FPGA CAD flows tend to be deterministic. For instance, in mesh topologies, deterministic routing algorithms such as dimension-order routing (DOR) \cite{seitz1988architecture} and zigzag routing \cite{badr1989optimal} serve as notable examples. DOR algorithms, such as XY-routing and YX-routing, route packets along one dimension until they align with the destination in that dimension, then switch to the second dimension. Conversely, zigzag routing alternates between two dimensions until aligning with the destination in one dimension, after which it proceeds along the other dimension. These deterministic algorithms, however, exhibit no path diversity, as the route between a source and destination is solely determined by their relative positions in the mesh. This lack of path diversity may lead to link oversubscription, as traffic flows are unnecessarily forced to share links.

\begin{figure}[tp]
     \centering
     \includegraphics[width=0.25\textwidth]{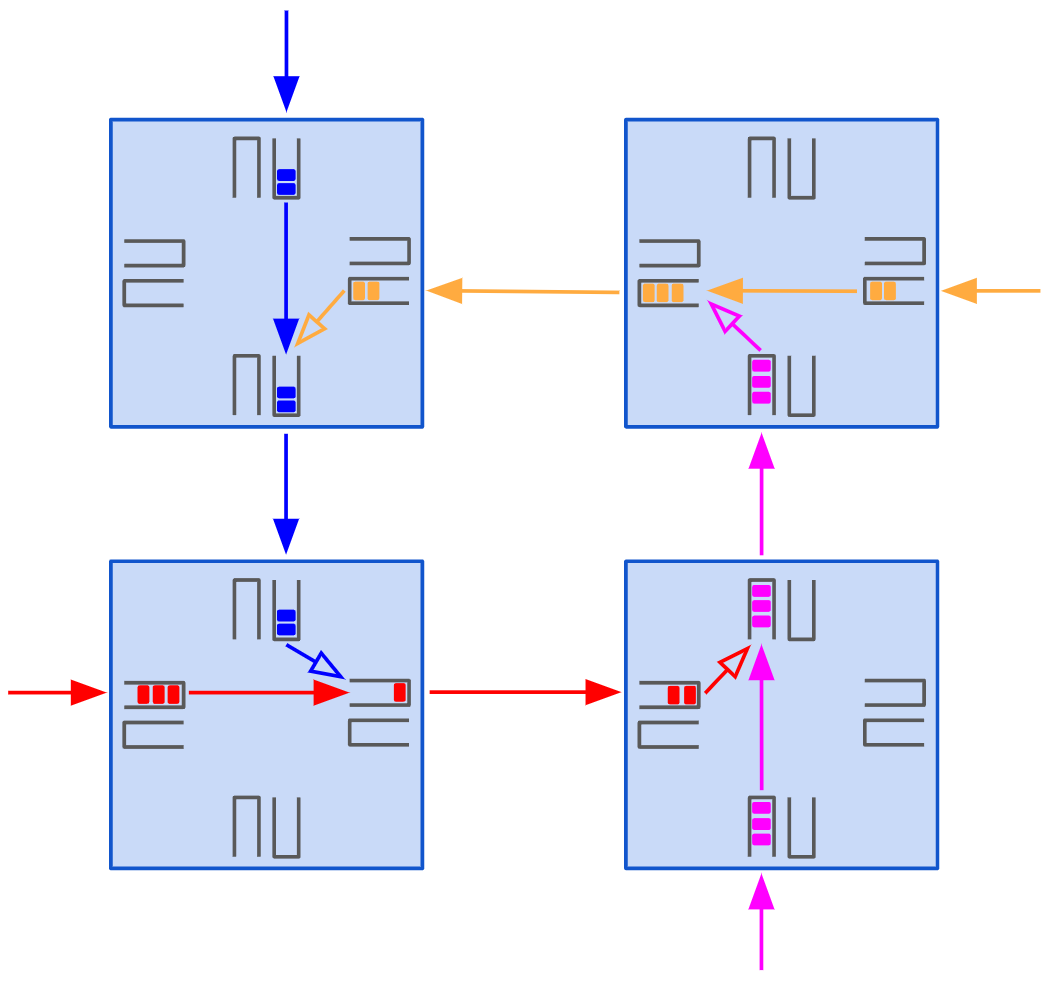}
    \caption{Four traffic flows (each in a different color) stuck in a deadlock due to cyclic buffer acquisition}
    \label{fig:deadlock}
\end{figure}
However, exploiting path diversity between NoC routers by routing traffic flows arbitrarily to avoid congestion is not always feasible due to the risk of deadlock. Fig.~\ref{fig:deadlock} illustrates a scenario where four traffic flow routes create a cyclic buffer acquisition dependency, preventing any from making progress. Therefore, it is essential to employ routing algorithms that can leverage path diversity while ensuring deadlock freedom.

In a mesh topology, a traffic flow can take eight possible turns. Glass et al. \cite{glass1994turn} demonstrated that cyclic acquisition of links can occur only if all four possible clockwise (or anti-clockwise) turns have been made (not necessarily by the same traffic flow). This implies that deadlocks can be prevented by prohibiting at least one clockwise turn and one anti-clockwise turn. XY-routing accomplishes this by forbidding four turns. However, deadlock-free routing algorithms that forbid only two turns, known as turn model algorithms, are also possible.

While turn model algorithms provide a higher degree of path diversity, they may lack path diversity for specific pairs of NoC routers due to the prohibition of certain turns. For instance, the west-first algorithm, which forbids all westward turns, fails to offer any path diversity when the destination router is situated to the west of the source router. To mitigate this limitation, Chiu \cite{chiu2000odd} proposed the odd-even routing algorithm. This algorithm ensures deadlock freedom without forbidding any turns; instead, it restricts the columns from which each turn can be taken. Specifically, it forbids two turns in odd columns and two other turns in even columns.

Fu et al. \cite{fu2011abacus} found that forbidding specific turns in each column can be too restrictive for some traffic patterns and introduced the abacus algorithm to increase flexibility. It associates two beads with each column; beads can move arbitrarily within their column. Routers above a bead forbid one turn, while routers below the bead forbid another turn. This algorithm has the same number of legal turns as odd-even routing, but moving beads across each column enables a more fine-grained distribution of legal turns than column-wise prohibition.

\section{Congestion-aware NoC Mapping}\label{sec:congestion_aware_noc_mapping}
\subsection{Current NoC Placement Support in VPR}
Srinivasan et al. \cite{srinivasan2023placement} enhanced the VPR architecture file syntax to include NoC topology and metric descriptions. With the enhanced VPR architecture file syntax, users can specify various NoC topologies, link bandwidth capacities, and latencies for links and routers. NoC latencies can be defined using either zero-load latencies or worst-case uncongested link and router latencies, which can be calculated analytically \cite{lang2021worst}. The netlist can include logical NoC routers that map to physical NoC routers in the FPGA grid. VPR's placement engine utilizes an RL-assisted simulated annealing optimizer \cite{mahmoudi2023respect} to gradually minimize placement cost by perturbing the locations of netlist blocks. For NoC traffic flow routing, VPR employs XY-routing, where traffic flow routes are determined solely by the location of their source and destination routers. Srinivasan et al. \cite{srinivasan2023placement} introduced new NoC-related cost terms to account for aggregate NoC bandwidth usage and traffic flow latency overruns. These new cost terms are multiplied by their respective weighting factors and added to the placement cost function, enabling the annealer to optimize total wirelength, critical path delay (CPD), and NoC-related terms simultaneously.

\subsection{Congestion Modeling}
For each NoC link $L_i$, we measure its utilized bandwidth, denoted as $LinkU(L_i)$, by accumulating the bandwidth of all traffic flows routed through $L_i$. Subsequently, we compare the link's bandwidth utilization with its available bandwidth, $LinkBW(L_i)$. The degree to which a link's bandwidth utilization exceeds its available bandwidth represents the level of congestion experienced by that link. This congestion measure is quantified by the following equation (\ref{eq:congestion_cost}):
\begin{equation}\label{eq:congestion_cost}
C_{cong} = \gamma \times \sum_{L_i \in \mathbb{L}} \max (0, LinkU(L_i) - LinkBW(L_i))
\end{equation}
In this equation, $C_{cong}$ represents the NoC congestion cost, computed as the sum of congestion experienced across all links. The accumulated NoC congestion is then scaled by a weighting factor ($\gamma$), which specifies its relative importance compared to other NoC cost terms.

The placement cost, as shown in Eq.~\eqref{eq:place_cost}, is modified to include a new term reflecting NoC congestion:
\begin{equation}\label{eq:place_cost}
C_{Total} = C_{netlist} + \omega \times (C_{bw} + C_{lat} + C_{cong}),
\end{equation}
where $C_{netlist}$ represents a weighted average of traditional wirelength and timing costs. The parameter $\omega$ determines the importance of NoC-related cost terms relative to $C_{netlist}$.

\subsection{Motivation for Path Diversity}\label{sec:motiv_diversity}
Optimizing NoC congestion by relocating logical routers during placement can mitigate congestion to some extent but often falls short of fully resolving it across a wide range of benchmarks. This is due to the XY-routing algorithm's inability to leverage path diversity between a given source and destination pair. To elaborate on this point, we adopt the concept of routing algorithm adaptiveness ($P_{\text{algorithm}}$) \cite{glass1992turn}, which refers to the number of distinct shortest-path routes available between a source NoC router located at $(x_{\text{src}}, y_{\text{src}})$ and a destination router at $(x_{\text{dst}}, y_{\text{dst}})$ within a mesh topology.

The XY-routing algorithm, by design, lacks path diversity, offering only a single route for each router pair, resulting in $P_{\text{XY}} = 1$. In contrast, a deadlock-oblivious routing algorithm that is able to take all turns can expose multiple paths as the distance between router pairs increases:
\begin{equation}\label{eq:p_deadlock_oblivious}
P_{\text{deadlock-oblivious}} = \frac{(|\Delta x| + |\Delta y|)!}{(|\Delta x|)! \times (|\Delta y|)!} \geq 1
\end{equation}

\begin{figure}[tp]
     \centering
     \begin{subfigure}[b]{0.12\textwidth}
         \centering
         \includegraphics[width=\textwidth]{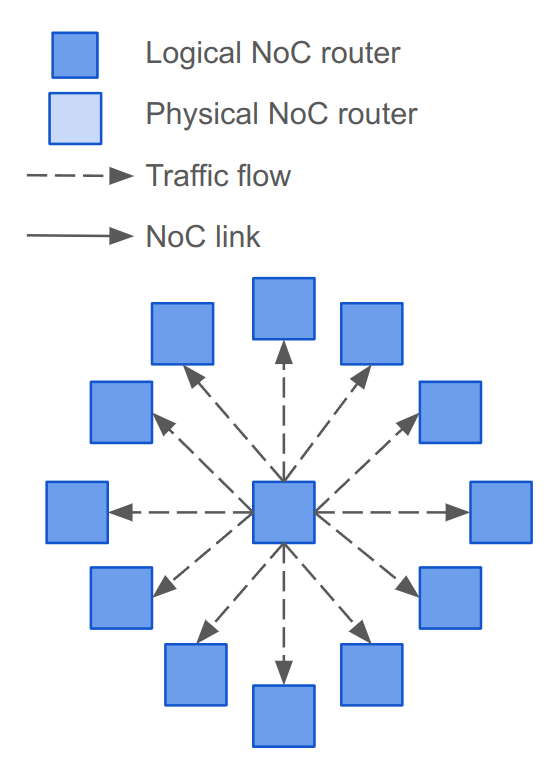}
         \caption{}
         \label{fig:star_logical}
     \end{subfigure}
     \begin{subfigure}[b]{0.16\textwidth}
         \centering
         \includegraphics[width=\textwidth]{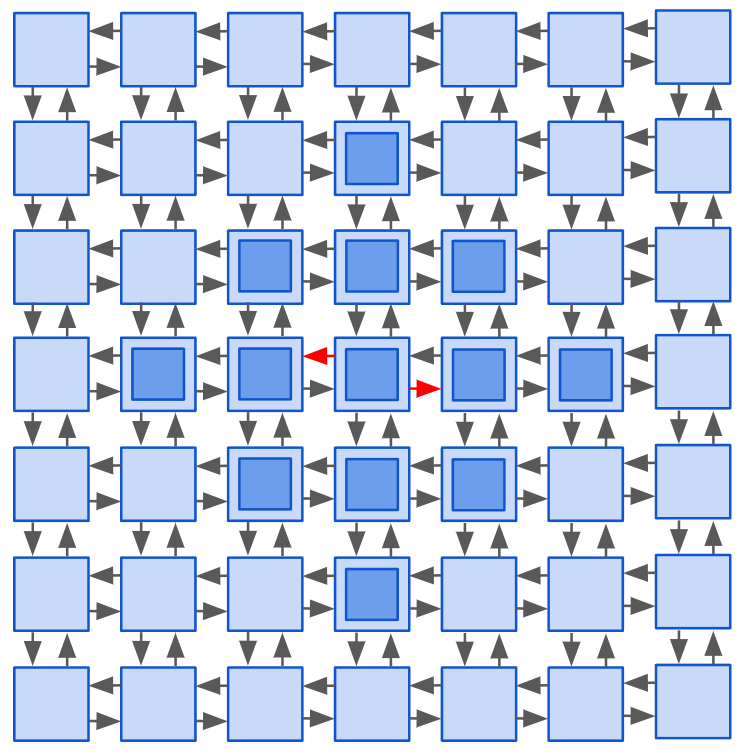}
         \caption{}
         \label{fig:star_cong_obliv}
     \end{subfigure}
     \begin{subfigure}[b]{0.16\textwidth}
         \centering
         \includegraphics[width=\textwidth]{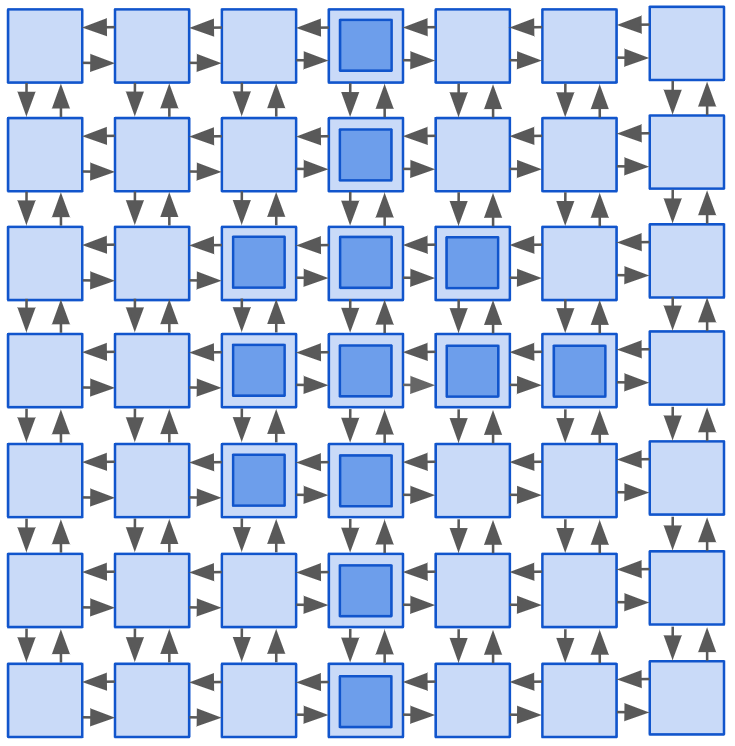}
         \caption{}
         \label{fig:star_cong_aware}
     \end{subfigure}
        \caption{Impact of congestion-awareness on NoC placement. (a)~Logical NoC routers and traffic flows between them. (b)~Congestion-oblivious mapping of NoC routers resulted in two congested links shown in red. (c)~Congestion-aware NoC placement resolved the congestion at the cost of higher aggregate bandwidth.}
        \label{fig:path_diversity_motiv}
\end{figure}

Fig.~\ref{fig:path_diversity_motiv} illustrates the trade-off inherent in congestion-aware NoC placement with XY-routing, particularly highlighting the balance between aggregate bandwidth and congestion avoidance. In Fig.~\ref{fig:star_logical}, traffic flows between 13 logical NoC routers are depicted, with each flow utilizing 33\% of the NoC link bandwidth. 
In Fig.~\ref{fig:star_cong_obliv}, we observe the outcome of a congestion-oblivious NoC placement strategy focused solely on optimizing aggregate bandwidth and latency. Here, the central logical router is placed in the middle of the chip, surrounded by other routers to minimize aggregate bandwidth. However, when utilizing XY-routing, the horizontal outgoing links of the central router become oversubscribed by 33\%. Contrastingly, Fig.~\ref{fig:star_cong_aware} showcases a congestion-aware placement approach that resolves congestion. This resolution, however, comes at the expense of higher aggregate bandwidth usage. It is worth noting that a more path-diverse routing algorithm could potentially address congestion within the placement depicted in Fig.~\ref{fig:star_cong_obliv} by leveraging path diversity.

\subsection{Integrating New NoC Routing Algorithms into VPR}
In \cite{srinivasan2023placement}, when the SA algorithm relocates a logical NoC router, all its associated traffic flows are rerouted using the XY-routing algorithm. We integrate three turn model-based algorithms introduced in \cite{glass1992turn} and the odd-even routing algorithm \cite{chiu2000odd} into VPR. Unlike the XY-routing algorithm, these algorithms expose path diversity and may offer multiple permissible directions at each intermediate NoC router along the traffic flow route.

Given that the annealer may attempt millions of moves, including swapping NoC routers tens of thousands of times, rerouting the associated traffic flows must be CPU-time efficient to avoid compromising the annealer's performance. To achieve this, we utilize the newly added NoC routing algorithms exclusively to generate minimal paths, prohibiting any detours. Consequently, the number of allowed directions is limited to a maximum of two. When the routing algorithm presents two possible direction choices at a router, we use a biased coin flip to randomly select one of them. This random decision is biased towards choosing the direction in which the distance to the destination is longer. For example, if the last router added to a partial route is located at $(x_\text{curr}, y_\text{curr}) = (3, 5)$ while the route is destined for $(x_\text{dst}, y_\text{dst}) = (7, 7)$, the random decision favors the X-direction twice as much as the Y-direction. This approach distributes the chance of selecting a shortest path more evenly among all possible paths between two NoC routers.

Disallowing detours simplifies the generation of traffic flow routes, making the evaluation of logical NoC router swaps relatively inexpensive for the SA algorithm. However, this limitation has a significant implication for the trade-off between aggregate bandwidth and congestion. As discussed in Section~\ref{sec:motiv_diversity}, the degree of adaptiveness of a deadlock-oblivious routing algorithm increases as routers are placed further apart. Consequently, the SA algorithm may opt to position NoC routers at greater distances to expose more path diversity. While this can help mitigate congestion, it comes at the expense of higher aggregate bandwidth utilization.

\subsection{NoC Routing as a SAT Problem}\label{sec:sat}
AMD has integrated a \textit{NoC Compiler} \cite{swarbrick2019versal} into their CAD flow to perform NoC placement and routing \textit{before} fabric resource placement. This NoC Compiler first places logical routers heuristically, and then formulates NoC routing as a SAT problem to avoid deadlocks, link bandwidth oversubscription, and conflicts in hierarchical routing tables. However, they do not reveal any details about the SAT formulation. We adopted an intrinsically different approach with the SA algorithm to co-optimize NoC placement, packet routing \textit{and} fabric resource placement. However, our approach may sometimes fail to find congestion-free minimal routes. In this case, we resort to a SAT formulation that is run after the annealer to improve NoC packet routing (without changing the placement) by enabling detours while avoiding deadlocks.

We assign a binary variable to each traffic flow and link pair $(T_i, L_j)$. When set, $(T_i, L_j)$ specifies that traffic flow $T_i$ is routed through link $L_j$. For each traffic flow $T_i$, we specify its source and destination routers by $T_i.src$ and $T_i.dst$. $L_{out}(R)$ and $L_{in}(R)$ determine the outgoing and incoming links of the NoC router $R$. For each traffic flow $T_i$, exactly one of the outgoing links of $T_i.src$ must be activated. We use the following constraints to enforce this restriction:

\begin{equation}\label{eq:outgoing_link_or}
\bigvee_{L_j \in L_{out}(T_i.src)}(T_i, L_j)
\end{equation}

\begin{equation}\label{eq:outgoing_link_and}
\bigwedge_{\substack{L_j, L_k \in L_{out}(T_i.src) \\ L_j \neq L_k}} \neg (T_i, L_j) \lor \neg (T_i, L_k)
\end{equation}

These constraints ensure that each traffic flow has exactly one activated outgoing link from its source. Similarly, we enforce that exactly one of the incoming links of $T_i.dst$ is set. Additionally, we need to ensure that the set of links selected for traffic flow $T_i$ forms a continuous route. For each router $R_j \neq T_i.src, T_i.dst$, we require at most one of its incoming flow-link variables to be set:

\begin{equation}\label{eq:at_most_one_incoming}
\bigwedge_{\substack{L_k, L_p \in L_{in}(R_j) \\ L_k \neq L_p}} \neg (T_i, L_k) \lor \neg (T_i, L_p)
\end{equation}

Similarly, we apply this constraint to the outgoing links of $R_j$. Additionally, when an incoming flow-link variable of $R_j$ is set, one of its outgoing flow-link variables must be set. We enforce this by applying the following constraint:

\begin{equation}\label{eq:exactly_one_incoming}
\forall L_k \in L_{in}(R_j), \neg (T_i, L_k) \lor \left[ \bigvee_{L_p \in L_{out}(R_j)}(T_i, L_p) \right]
\end{equation}

To guarantee deadlock freedom, we rely on turn model or odd-even routing algorithms to specify all illegal turns. Each illegal turn is determined by two consecutive NoC links that must not be traversed one after another. These constraints are translated to Boolean constraints and added to our SAT formulation.

To detect congestion, we introduce an auxiliary Boolean variable $C_j$ for each link $L_j$. For each traffic flow $T_i$ and link $L_j$, we quantize their bandwidth to $T_i.bw$ and $L_j.bw$ with a constant resolution. The following constraint sets $C_j$ when the traffic flow bandwidth traveling through $L_j$ exceeds $L_j.bw$:

\begin{equation}\label{eq:congestion_var}
\sum_{T_i \in \mathbb{T}} (T_i, L_j) \times T_i.bw > L_j.bw \rightarrow C_j
\end{equation}

The linear expression in Eq.~\eqref{eq:congestion_var} can be converted to a set of Boolean constraints through binary encoding \cite{abio2014encoding}. Moreover, we translate each traffic flow latency constraint to a maximum number of links that the traffic flow can traverse before violating the latency constraint and encode it into $T_i.lat$. We introduce auxiliary variable $V_i$ to indicate whether the latency constraint of traffic flow $T_i$ is violated:

\begin{equation}\label{eq:latency_violation_var}
\sum_{L_j \in \mathbb{L}} (T_i, L_j) > T_i.lat \rightarrow V_i
\end{equation}

We define the following objective function for the SAT solver:

\begin{equation}\label{eq:sat_obj}
\sum_{\substack{(T_i, L_j) \ T_i \in \mathbb{T}, L_j \in \mathbb{L}}}(T_i,L_j)\times T_i.bw + \alpha_q\sum_{V_i \in \mathbb{V}} V_i + \gamma_q\sum_{C_j \in \mathbb{C}} C_j, 
\end{equation}
where $\alpha_q$ and $\gamma_q$ are quantized weighting factors for latency overrun and congestion cost terms. Using this objective function, the SAT solver generates legal, continuous, deadlock-free routes while minimizing congestion, traffic flow latency overrun, and aggregate bandwidth. One can prioritize different NoC metrics by varying the objective function weighting factors.

\section{NoC-aware Packing and Placement Optimizations}\label{sec:pack_place_opt}

\subsection{NoC-biased Centroid Move}\label{sec:directed_move}
VPR employs a simulated annealing (SA) optimizer to iteratively enhance the total wirelength and critical path delay (CPD) of an initial constructive placement by perturbing the locations of movable blocks. However, except for blocks that must maintain a fixed relative placement (e.g., due to carry chains and DSP block cascades), the SA algorithm restricts perturbations, or moves, to involve at most two blocks. As the temperature decreases and the placement stabilizes during SA optimization, moves tend to localize. This localized behavior can present challenges when optimizing placements in NoC-enhanced FPGAs.

\begin{figure}[tp]
     \centering
     \begin{subfigure}[tp]{0.14\textwidth}
         \centering
         \includegraphics[width=\textwidth]{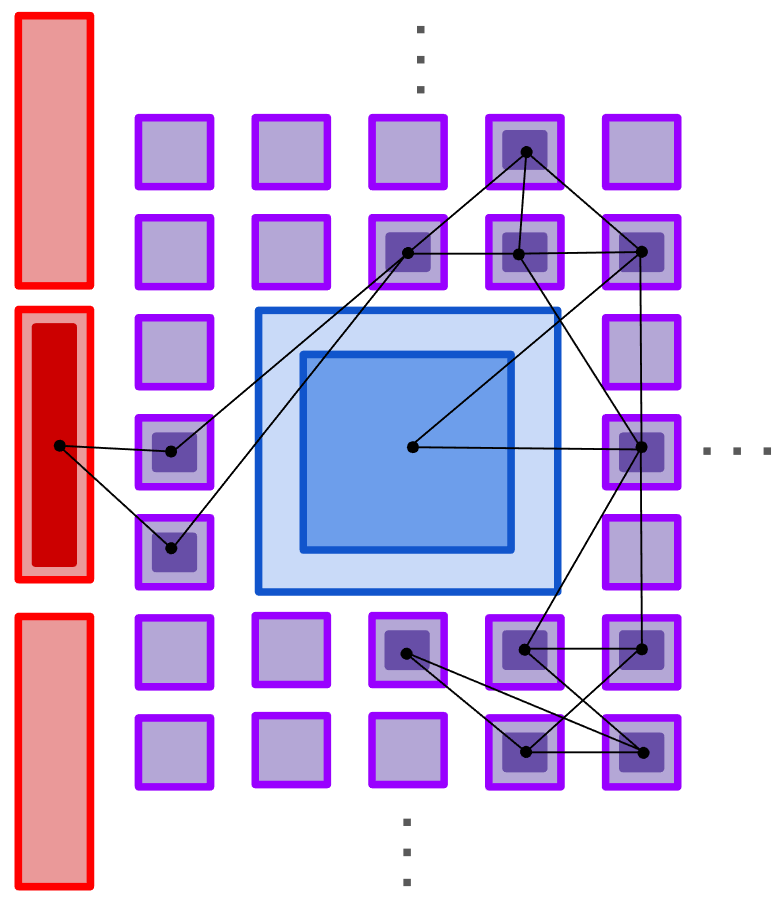}
         \caption{}
         \label{fig:noc_pre_swap}
     \end{subfigure}
     \hfill
     \begin{subfigure}[tp]{0.28\textwidth}
         \centering
         \includegraphics[width=\textwidth]{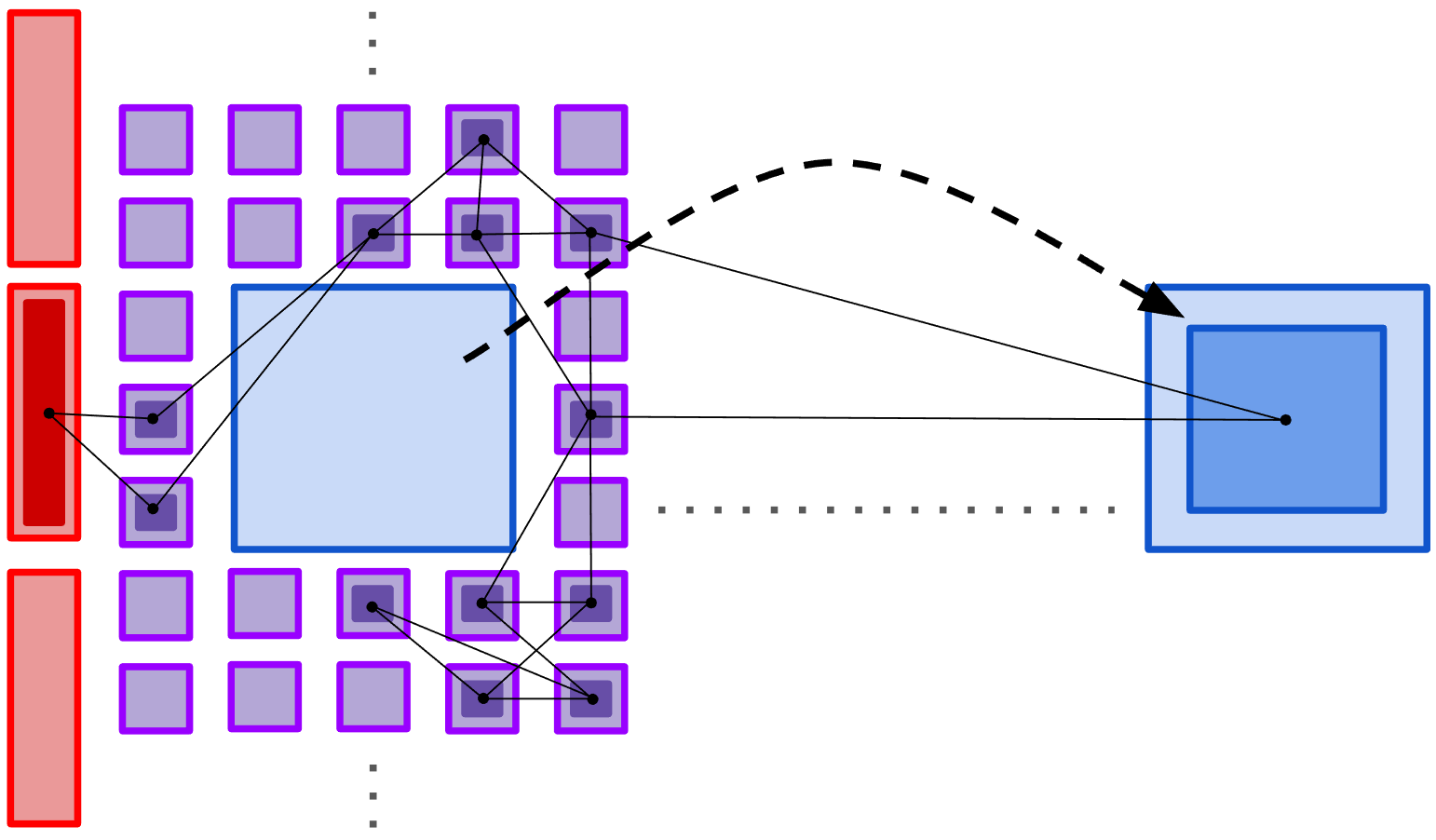}
         \caption{}
         \label{fig:noc_post_swap}
     \end{subfigure}
        \caption{Inter-block connectivity before and after a NoC swap}
        \label{fig:noc_swap}
\end{figure}

For instance, Fig.~\ref{fig:noc_swap} illustrates a scenario where the annealer successfully swaps a NoC router to a different location to improve NoC-related cost terms. However, other blocks fail to follow suit and remain stagnant in their original positions, as depicted in Fig.~\ref{fig:noc_post_swap}. This stagnation arises because these blocks are heavily interconnected, with only a few of them directly linked to the router. Consequently, when a NoC-connected block attempts to move closer to the router, the increased wirelength cost of inter-block nets outweighs the reduction in cost for the NoC-connected nets. As a result, most such swaps are rejected by the annealing process, necessitating a large number of moves with a positive cost difference (hill-climbing) to allow the group of blocks to converge towards their corresponding NoC router.

VPR's annealer uses a variety of perturbation strategies, or move types. Building upon the centroid move introduced in \cite{elgammal2021rlplace}, we propose a new type of move: the NoC-biased centroid move. In this move, each block is assigned to a NoC group. A NoC group $G_b$ associated with movable block $b$ encompasses all blocks reachable from a specific NoC router solely by traversing low fanout nets, i.e., nets with a fanout of less than 10. It should be noted that a NoC group may contain multiple NoC routers if they are reachable from each other through low fanout fabric nets. These NoC groups are formed as a pre-processing step before the placement stage via a breadth-first search of the design netlist in which high fanout nets are not traversed; high fanout nets are ignored as they indicate a weaker coupling between fabric resources.

The NoC-biased centroid move initially computes a target location $(X_c, Y_c)$ for block $b$ as the centroid of the blocks to which it is connected, similar to the bounded centroid move of \cite{elgamma2020learn}. Subsequently, the computed centroid location is slightly adjusted towards the NoC routers in $G_b$ by calculating a weighted average between the computed centroid location, $(X_c, Y_c)$, and the locations of the routers in the group, $(X_R, Y_R)$. The following equation shows how the adjusted location along the horizontal axis is calculated:
\begin{equation}\label{eq:noc_centroid_move}
X = (1 - W_{NoC}) \cdot X_c + \frac{W_{NoC}}{|G_b \cap \mathbb{R} |}  \sum_{R \in G_b\ \cap \mathbb{R}} X_R,
\end{equation}
where $\mathbb{R}$ denotes the set of all NoC routers and $W_{NoC}$ is a weighting factor specifying how much the block location should be adjusted towards its corresponding NoC router(s). The location along the vertical axis is computed similarly.

\subsection{NoC-aware Packing}\label{sec:packing}
VPR employs a seed-based packing algorithm \cite{luu2011architecture, luu2014towards} designed to capture low fanout connectivity within clustered blocks, thus minimizing demand on inter-cluster routing wires. However, in order to prevent under-utilization of primitives within a clustered block, the packing algorithm may resort to high fanout nets to infer logical connectivity for selecting candidate primitives for growing the cluster. In designs targeting NoC-enhanced FPGAs, this approach can lead to issues, especially since modules often communicate through the NoC, with high fanout nets such as reset and clock enable signals being the only nets connecting them through the fabric.

\begin{figure}[tp]
     \centering
     \begin{subfigure}[tp]{0.23\textwidth}
         \centering
         \includegraphics[width=\textwidth]{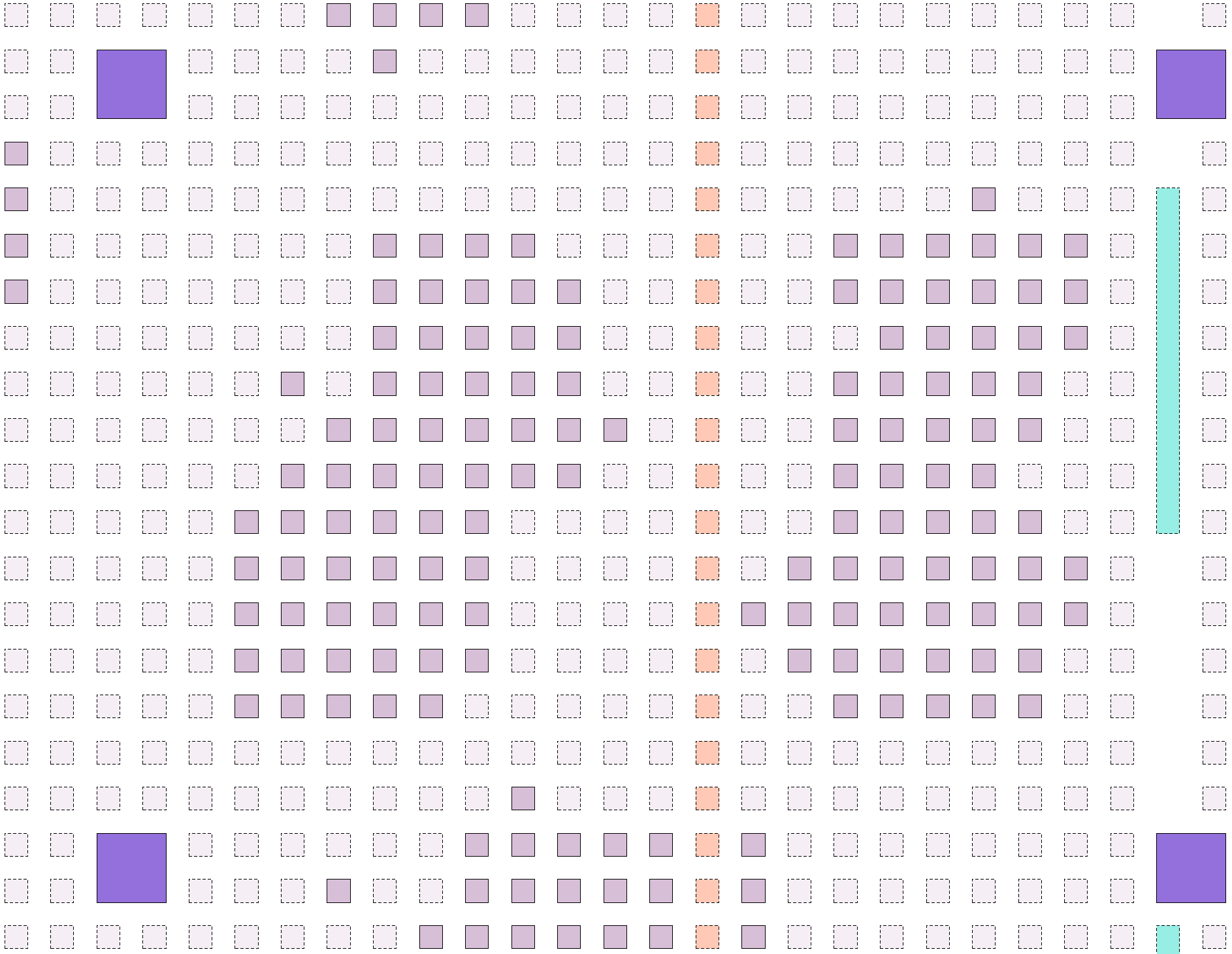}
         \caption{}
         \label{fig:noc_spread_place}
     \end{subfigure}
     \begin{subfigure}[tp]{0.23\textwidth}
         \centering
         \includegraphics[width=\textwidth]{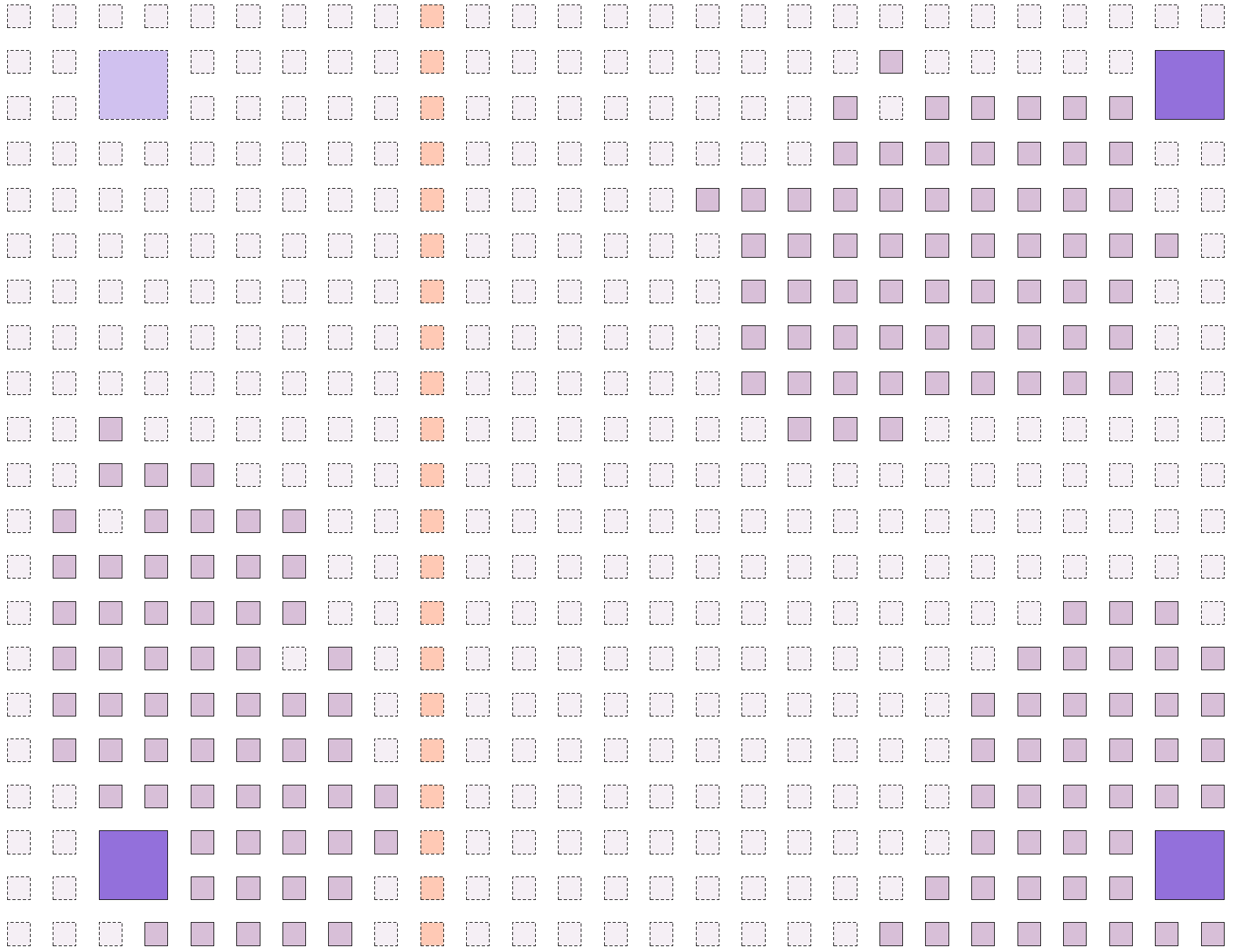}
         \caption{}
         \label{fig:noc_local_place}
     \end{subfigure}
     \hfill
        \caption{(a)~Clustered blocks are spread between NoC routers as some of them contain primitives from multiple NoC-attached modules. (b)~Localized placement of clustered blocks around their corresponding NoC router after NoC-aware packing optimization.}
        \label{fig:noc_place}
\end{figure}

VPR's packing algorithm may attempt to pack primitives from separate NoC-attached modules based on their connectivity through high fanout nets. Consequently, other primitives connected through low fanout nets to either of these two primitives might be added to the cluster based on their low fanout connectivity to the growing cluster. During the placement stage, clustered blocks containing primitives from two NoC-attached modules cannot be placed near either NoC router. Instead, they are stretched across the space between the two NoC routers as illustrated in Fig.~\ref{fig:noc_spread_place}.

To mitigate this issue, we employ a breadth-first search (BFS) starting from NoC routers in the primitive netlist to identify all blocks reachable from each NoC router through low fanout nets. This step divides the netlist graph into several connected components. During the packing stage, we prohibit the packing of primitives from different connected components. As a result, each clustered block includes primitives reachable from at most a single NoC router. With this optimization, the placement engine can place clusters close to their corresponding NoC router as shown in Fig.~\ref{fig:noc_local_place}.

\section{Experimental Results}\label{sec:results}
\begin{figure}[tp]
     \centering
     \begin{subfigure}[tp]{0.1\textwidth}
         \centering
         \includegraphics[width=\textwidth]{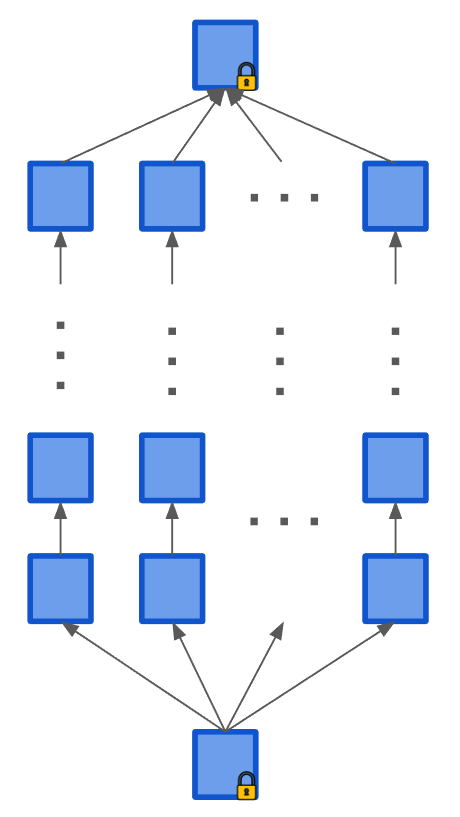}
         \caption{}
         \label{fig:genome_seq}
     \end{subfigure}
     \hfill
     \begin{subfigure}[tp]{0.1\textwidth}
         \centering
         \includegraphics[width=\textwidth]{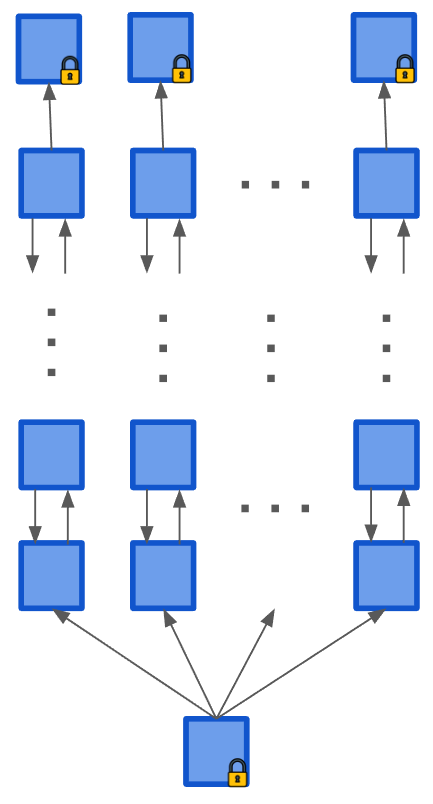}
         \caption{}
         \label{fig:page_rank}
     \end{subfigure}
    \hfill
     \begin{subfigure}[tp]{0.11\textwidth}
         \centering
         \includegraphics[width=\textwidth]{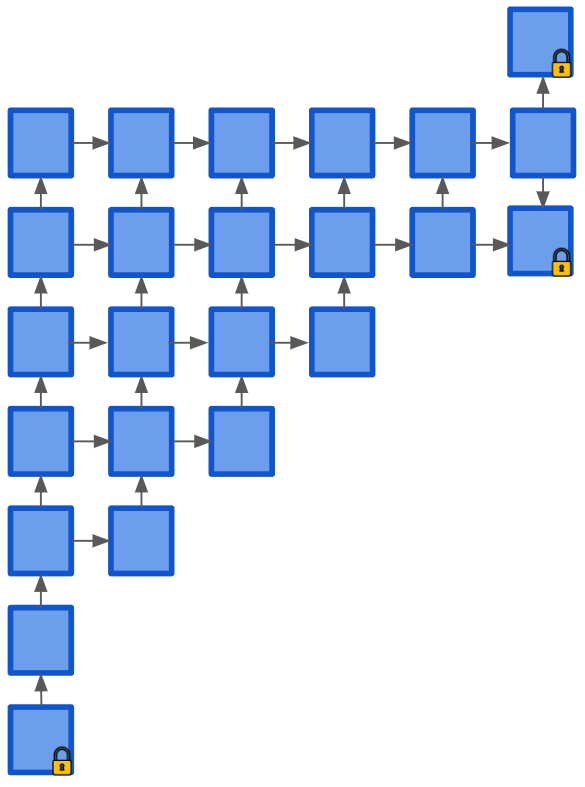}
         \caption{}
         \label{fig:gaussian_elim}
     \end{subfigure}
     \hfill
     \begin{subfigure}[tp]{0.095\textwidth}
         \centering
         \includegraphics[width=\textwidth]{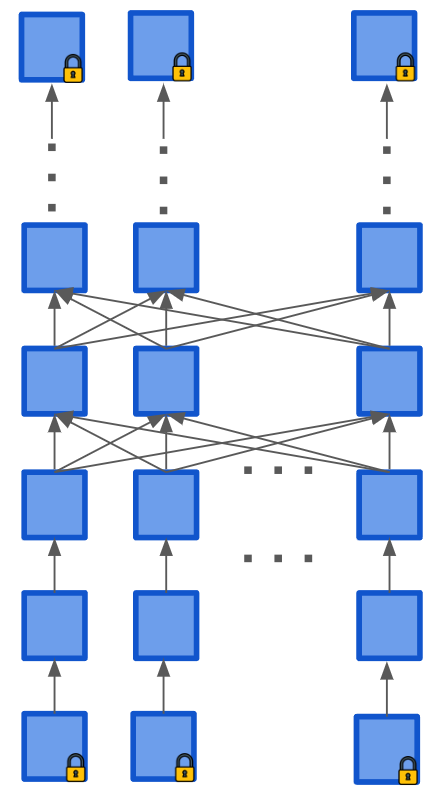}
         \caption{}
         \label{fig:bucket_sort}
     \end{subfigure}
     
        \caption{Traffic flow patterns in new synthetic benchmarks}
        \label{fig:new_syn_benchmarks}
\end{figure}

In this section, we evaluate the different optimizations proposed in previous sections using a synthetic benchmark suite consisting of 29 circuits. We utilize the \textit{star} benchmarks introduced in \cite{srinivasan2023placement}, excluding the \textit{2D nearest neighbor} and \textit{1D chain} benchmarks, as finding congestion-free placement and packet routing solutions for these benchmarks is trivial. Instead, we introduce new synthetic benchmarks that mimic the high-bandwidth traffic flows generated by hardened memory controllers or high-speed I/O interfaces by locking down some NoC routers to locations at the periphery of the FPGA device. Fig.~\ref{fig:new_syn_benchmarks} illustrates the traffic flows in these new synthetic benchmarks. The following describes our new synthetic benchmarks:

\begin{itemize}
    \item \textit{Genome sequencing}: Inspired by the architecture proposed in \cite{guo2019hardware}, we lock down two logical NoC routers on opposite sides of an FPGA device to emulate hardened memory controllers. This benchmark consists of nine processing element (PE) lines where multiple NoC-attached PEs are connected in a chain. The first fixed NoC router transmits traffic flows to the first PE in each line, while the last PE in each line transmits a traffic flow to the other fixed NoC router.
    \item \textit{Page Rank}: Similar to the genome sequencing benchmark, a logical NoC router, which distributes traffic flows to nine PE lines, is locked down at the bottom of the device. Inter-PE traffic flows in each line flow in both directions. At the end of each line, a logical NoC router is locked down at the top of the device to emulate an external memory interface. This traffic flow pattern was inspired by the architecture introduced in \cite{chi2021extending}.
    \item \textit{Gaussian elimination}: Inspired by the triangular topology introduced in \cite{wang2021autosa}, we lock down NoC routers at opposite corners of the device, while NoC-attached PEs transmit traffic flows in a triangular pattern. When the triangle's dimensions are smaller than the NoC, a congestion-free mapping can be easily found. However, in our benchmarks, perpendicular sides of the triangle have 11 NoC routers while the FPGA being targeted contains a $10\times10$ mesh of physical NoC routers, making a congestion-free solution challenging.
    \item \textit{Bucket sort}: Based on the architecture proposed in \cite{samardzic2020bonsai}, we lock down 8 NoC routers at two opposite sides of the device. Each PE line is a chain of NoC-attached PEs starting from one of the bottom NoC routers and ending at one of the top routers. There are two fully connected layers that redistribute the data between PE lines, complicating a congestion-free mapping.
\end{itemize}

In our experiments with synthetic benchmarks, we use an approximation of the Stratix-IV architecture \cite{murray2015timing} with a $10\times10$ mesh NoC embedded into the fabric.

\subsection{Aggregate Bandwidth-Congestion Trade-off}
In Section~\ref{sec:motiv_diversity}, we discussed how path diversity can aid congestion resolution. However, we observed that exposing more path diversity, i.e., having a larger $P_{algorithm}$, requires placing NoC routers farther from each other, thus increasing aggregate bandwidth. To evaluate the trade-off between congestion resolution and aggregate bandwidth utilization, we conducted experiments with different NoC routing algorithms and varying values of $\gamma$ using 29 synthetic benchmarks, each run with three different random placement seeds to reduce CAD noise. Table~\ref{tab:bw_congestion_tradeoff} illustrates various QoR metrics and placement runtime with respect to the placement engine without congestion modeling.

As Table~\ref{tab:bw_congestion_tradeoff} shows, NoC-congestion-aware placement is effective as all routing algorithms see a large decrease (from 2.5$\times$ to 20$\times$) in congestion when $\gamma$ is non-zero. Congestion tends to decrease as $\gamma$ increases. Path-diverse routing algorithms are more effective than the XY-routing algorithm in resolving congestion, and have a synergistic interaction with NoC-congestion-aware placement -- they always reduce congestion, but the reduction is much larger with NoC-congestion-aware placement. For instance, the XY-routing algorithm fails to resolve congestion in 79 cases (at the best value of $\gamma$), while odd-even routing fails to fully resolve congestion in only 9 cases. Another observable trend in Table~\ref{tab:bw_congestion_tradeoff} is the increase (4\% to 14\%, depending on the packet routing algorithm and $\gamma$ value) in aggregate bandwidth as the weighting of NoC congestion in the placement cost increases. This occurs because the annealer tends to place NoC routers farther from each other to expose more path diversity to the underlying NoC routing algorithm.

To demonstrate the trade-off between aggregate bandwidth and congestion resolution, we computed the aggregate bandwidth-congestion product for each experiment and included it in Table~\ref{tab:bw_congestion_tradeoff}. Path-diverse routing algorithms strike a better balance between these two competing metrics than the XY-routing algorithm, with odd-even and west-first algorithms outperforming other algorithms. However, the west-first algorithm violates a large number of latency constraints as it cannot offer diverse paths for half of the source-destination pairs. 

In Table~\ref{tab:bw_congestion_tradeoff}, for the XY-routing algorithm, the $\gamma=\text{off}$ experiment does not include the congestion modeling implementation, while the $\gamma=0$ experiment models NoC congestion but weights its cost as 0. Congestion modeling increases the complexity of the cost change calculation for NoC router swaps, but since NoC router swaps make up only a small fraction of all attempted swaps, the extra complexity has negligible impact on placement runtime. 

The increased runtime in path-diverse routing algorithms compared to XY-routing can be attributed to more attempted NoC swaps. This is because the RL agent learns that NoC swaps are more effective in reducing NoC congestion and thereby the placement cost when a path-diverse routing algorithm is used. When $\gamma$ increases, congestion cost contributes more to the total placement cost, increasing the number of attempted NoC swaps. Moreover, path-diverse routing algorithms are more complex than XY-routing, causing each NoC swap to take more time to route associated traffic flows.

The final two columns in Table~\ref{tab:bw_congestion_tradeoff} show that the programmable interconnect usage and the critical path delay of the designs are essentially unaffected by NoC-congestion-aware placement.

In summary, the odd-even routing algorithm offers the best trade-off between aggregate bandwidth utilization and congestion resolution while meeting more latency constraints than other path-diverse NoC routing algorithms.

\begin{table*}[tp]
    \centering
    \resizebox{\textwidth}{!}{
    \begin{tabular}{|c|c|c|c|c|c|c|c|c|c|}
    \hline
    Algorithm & $\gamma$ & Agg. BW & Congestion & Agg. BW $\times$ Congestion & Violated Lat. Cons. & Congested & Place Time & Est. WL & Est. CPD \\
    \hline
    \multirow{4}{*}{XY} & off & 1.00 & 1.000 & 1.000 & 4 & 87 & 1.00 & 1.000 & 1.000 \\
     & 0.0 & 1.00 & 1.066 & 1.070 & 3 & 86 & 1.01 & 1.002 & 1.000 \\
     & 0.25 & 1.04 & 0.378 & 0.391 & 3 & 79 & 1.00 & 1.005 & 1.007 \\
     & 5.0 & 1.08 & 0.356 & 0.383 & 16 & 80 & 1.02 & 1.010 & 1.001 \\
    \hline
    \multirow{3}{*}{West-first} & 0.0 & 1.00 & 0.938 & 0.938 & 3 & 86 & 1.04 & 1.002 & 1.000 \\
     & 0.25 & 1.08 & 0.128 & 0.138 & 99 & 13 & 1.04 & 1.009 & 0.997 \\
     & 5.0 & 1.12 & 0.067 & 0.074 & 112 & 10 & 1.07 & 1.013 & 1.003 \\
    \hline
    \multirow{3}{*}{North-last} & 0.0 & 1.00 & 0.889 & 0.889 & 3 & 86 & 1.03 & 1.002 & 1.000 \\
     & 0.25 & 1.07 & 0.177 & 0.189 & 8 & 58 & 1.04 & 1.003 & 0.998 \\
     & 5.0 & 1.14 & 0.154 & 0.175 & 93 & 39 & 1.05 & 1.004 & 1.005 \\
    \hline
    \multirow{3}{*}{Negative-first} & 0.0 & 1.00 & 0.912 & 0.912 & 3 & 87 & 1.04 & 1.002 & 1.000 \\
     & 0.25 & 1.08 & 0.121 & 0.131 & 66 & 16 & 1.05 & 0.993 & 0.998 \\
     & 5.0 & 1.13 & 0.089 & 0.101 & 124 & 9 & 1.07 & 0.989 & 0.999 \\
    \hline
    \multirow{4}{*}{Odd-even} & 0.0 & 1.00 & 0.840 & 0.836 & 3 & 87 & 1.09 & 1.002 & 1.000 \\
     & 0.25 & 1.04 & 0.093 & 0.097 & 9 & 14 & 1.10 & 1.004 & 0.999 \\
     & 0.25$^*$ & 1.05 & 0.049 & 0.051 & 10 & 10 & 1.11 & 1.004 & 0.999 \\
     & 5.0 & 1.09 & 0.072 & 0.079 & 21 & 9 & 1.12 & 0.998 & 1.000 \\
    \hline
    \end{tabular}
    }
    \caption{QoR metrics and placement runtime for different NoC routing algorithms and varying values of $\gamma$. The benchmark suite consists of 29 designs. Each experiment runs the CAD flow with 3 different seeds, resulting in 87 runs for each experiment. Reported aggregate bandwidth, placement time, estimated wirelength, and CPD are the geometric mean values over 87 runs. Congestion cost can become zero, so we use the arithmetic mean to report the average congestion. \\ $^*$ Indicates results after invoking the SAT solver}
    \label{tab:bw_congestion_tradeoff}
\end{table*}

\subsection{SAT NoC Routing}
\begin{table}[tp]
    \centering
    \begin{tabular}{|c|c|c|c|}
    \hline
    Metric & Before SAT-routing & After SAT-routing & Diff.  \\
    \hline
    Agg. BW & $5.66e7$ & $5.43e7$ & +4.2\% \\
    Agg. Lat. & $2.65e{-7}$ & $2.75e{-7}$ & +3.8\% \\
    Congestion & 1.419 & 0.747 & -47.3\% \\
    \# of Congested & 14 & 10 & -28.6\% \\
    Viol. Lat. Cons. & 1 & 2 & +1 \\
    
    \hline
    \end{tabular}
    \caption{QoR metric differences after invoking the CP-SAT solver on 14 congested cases with $\gamma=0.25$}
    \label{tab:sat_results}
\end{table}

We investigated if our SAT formulation for NoC routing could further improve congestion by running it on the output of the best-performing placement plus NoC routing algorithm in Table~\ref{tab:bw_congestion_tradeoff}, which is the odd-even routing algorithm with $\gamma=0.25$.
We implemented the SAT formulation proposed in Section~\ref{sec:sat} using the CP-SAT solver \cite{cpsatlp}, which is part of Google's OR-Tools \cite{ortools}. We limited the number of parallel workers in the CP-SAT solver to 16. %
To verify the legality of the generated routes, we again employed the CP-SAT solver, ensuring their continuity and deadlock-freedom. For the deadlock-freedom check, we constructed the channel dependency graph (CDG) \cite{dally1987deadlock} and examined it for cycles -- the absence of cycles in the CDG guarantees deadlock-freedom.

As seen in Table~\ref{tab:bw_congestion_tradeoff}, when $\gamma=0.25$, the odd-even routing algorithm fails to fully resolve congestion in 14 out of 87 experiments on synthetic benchmarks. Table~\ref{tab:sat_results} demonstrates how SAT-routing can further resolve congestion by detouring some traffic flows to avoid congested links; it resolves congestion in 4 additional benchmarks.

The SAT-router is not limited to minimal routes and can take detours to reduce congestion at the expense of higher aggregate bandwidth and latency. In the 14 cases where SAT-routing was applied, the aggregate bandwidth and latency increased by 4.3\% and 3.9\%, respectively. Taking detours helped the SAT solver reduce congestion by 47.3\% and fully resolve congestion in four cases. In one case, the SAT-router decided to violate a traffic flow latency constraint to resolve congestion. In the worst case, the SAT-router increased the total placement (including NoC packet routing) runtime by 11\%, while on average over 14 congested cases, the placement runtime increased only by 4.2\%.

As seen in Table~\ref{tab:bw_congestion_tradeoff}, when considering the entire benchmark suite, odd-even routing with $\gamma=0.25$ and SAT-routing strike a better balance between aggregate bandwidth and congestion than odd-even routing alone with a higher congestion weight of $\gamma=5.0$. This suggests that the placement engine can use a path-diverse NoC routing algorithm with small values of $\gamma$ to avoid severe unresolvable congestion, while post-placement SAT-routing resolves most minor congestions by taking detours.

\subsection{NoC-biased Centroid Move}
To evaluate the impact of the new NoC-biased centroid move on QoR metrics, we selected the odd-even routing algorithm and set $\gamma=0.25$. We enabled the new move type along with the packing optimization introduced in Section~\ref{sec:pack_place_opt} and conducted experiments with varying values of $W_{NoC}$. As shown in Table~\ref{tab:dir_results}, on average over the synthetic benchmarks the total wirelength is reduced by 8.8\% with minimal impact on the critical path delay. When the packing optimization is disabled, the NoC-biased move can only improve the wirelength by 3.3\%, highlighting the importance of NoC-awareness in the packing stage. Forbidding primitives attached to different NoC routers from being packed together slightly increases the number of clustered blocks by 1.2\%. 

We also tested the packing optimization and NoC-biased move on the multi-layer perceptron (MLP) complete designs from \cite{srinivasan2023placement}. As shown in Table~\ref{tab:dir_results}, these optimizations result in a reduction of wirelength by 2.3\% and CPD by 2.1\%, with only a slight increase of 0.1\% in the number of clustered blocks.

\section{Conclusion}\label{sec:conclusion}
In this paper, we presented several approaches to reduce congestion in FPGAs with hardened NoCs, while continuing to optimize both other NoC metrics and the key fabric metrics of programmable routing wirelength and critical path delay. All these algorithms are implemented in the VTR open-source FPGA CAD tool so other researchers can build upon them. Firstly, we modeled NoC congestion and incorporated it as a new cost term in the placement cost function, resulting in a 62.3\% reduction in NoC congestion at the expense of a 4\% increase in aggregate bandwidth. Secondly, we integrated path-diverse NoC routing algorithms into the CAD flow to evenly distribute traffic flows across NoC links and mitigate congestion. Together, these algorithms achieved a 90.3\% reduction in NoC congestion with only a 4\% increase in aggregate bandwidth. Additionally, to enable traffic flow detours, we formulated NoC routing as a SAT problem; when combined with the earlier two algorithms this results in a 95.1\% reduction in congestion compared to the baseline, with a 5\% higher aggregate bandwidth.

We also introduced a packing optimization and a new directed move for the placement stage, aiming to enhance traditional QoR metrics in designs targeting NoC-enhanced FPGAs. These optimizations led to a 2.3\% reduction in wirelength and a 2.1\% reduction in critical path delay in complete MLP benchmarks.

While the SAT formulation, placement moves, and placement cost function support arbitrary NoC topologies, the deadlock-free NoC routing algorithms we employ during placement are currently specific to mesh and torus NoC topologies. In our future work we plan to extend congestion optimization to a wider range of NoC topologies, and develop a broader range of benchmarks for NoC-enhanced FPGAs.

\begin{table}[tp]
    \centering
    \begin{tabular}{|c|c|c|c|c|c|}
    \hline
    \multicolumn{6}{|c|}{Synthetic Benchmarks} \\ 
    \hline
    $W_{NoC}$ & Packing Opt. & Packed Blocks & WL & CPD & Swaps  \\
    \hline
    \xmark & \xmark  & 1.000 & 1.000 & 1.000 & 1.000 \\
    0.2    & \cmark  & 1.012 & 0.925 & 1.000 & 1.016 \\
    0.4    & \cmark  & 1.012 & 0.912 & 0.998 & 1.024 \\
    0.6    & \cmark  & 1.012 & 0.916 & 1.004 & 1.026 \\
    0.8    & \cmark  & 1.012 & 0.921 & 1.003 & 1.029 \\
    1.0    & \cmark  & 1.012 & 0.929 & 1.001 & 1.025 \\
    0.4    & \xmark  & 1.012 & 0.967 & 1.000 & 0.994 \\
    \xmark & \xmark  & 1.000 & 1.007 & 1.001 & 1.031 \\
    \hline
    \multicolumn{6}{|c|}{MLP Benchmarks} \\ 
    \hline
    \xmark & \xmark  & 1.000 & 1.000 & 1.000 & 1.000 \\
    0.4    & \cmark  & 1.001 & 0.977 & 0.979 & 1.000 \\
    \hline
    \end{tabular}
    \caption{QoR metrics for different values of $W_{NoC}$ when $\gamma=0.25$ with the odd-even routing algorithm}
    \label{tab:dir_results}
\end{table}

\section*{Acknowledgment}
This work was supported by the Intel/VMware Crossroads 3D-FPGA Academic Research Centre, NSERC and QuickLogic.

\bibliographystyle{ieeetr}
\bibliography{ref}

\vspace{12pt}

\end{document}